%% file: negative_parity.tex
\documentclass[final,3p,times,twocolumn]{elsarticle}
                                   
\usepackage{graphicx}
\usepackage{amssymb}
\usepackage{amsmath}
\usepackage{hyperref} 
\usepackage{longtable}
\biboptions{sort&compress}

\usepackage{xcolor}

\journal{Physics Letters B}

\newcommand{\preprint}{
 \setlength{\unitlength}{1mm}{\hbox{\begin{picture}(0,0)
      \put(160,10){\mbox{\footnotesize%
        ADP-10-15/T711}}\end{picture}}}}

\begin{document}

\begin{frontmatter}

\title{\preprint 
       Ordering of Spin-$\frac{1}{2}$ Excitations of the Nucleon in Lattice QCD}

\author[adl,raj]{M.S. Mahbub}
\author[adl]{Waseem Kamleh}
\author[adl]{Derek B. Leinweber}
\author[adl,cyp]{Alan \'O Cais}
\author[adl]{Anthony G. Williams}
\address[adl]{Special Research Centre for the Subatomic Structure of Matter, Adelaide, South Australia 5005, Australia, \\ and Department of Physics, University of Adelaide, South Australia 5005, Australia.}
\address[raj]{Department of Physics, Rajshahi University, Rajshahi 6205, Bangladesh.}
\address[cyp]{Cyprus Institute, Guy Ourisson Builiding, Athalassa Campus, PO Box 27456, 1645 Nicosia, Cyprus.}

\begin{abstract}

We present results for the negative parity low-lying state of the
nucleon, $N{\frac{1}{2}}^{-}$ (1535 MeV) ${\rm S}_{11}$, from a variational
analysis method. The analysis is performed
in quenched QCD with the FLIC fermion
action. The principal focus of this paper
is to explore the level ordering between the
Roper (${\rm{P}}_{11}$) and the
negative parity ground (${\rm{S}}_{11}$) states of the
nucleon. Evidence of the physical level ordering is observed at light quark
masses. A wide variety of smeared-smeared correlation functions are
used to construct correlation matrices. A comprehensive
correlation matrix analysis is performed to ensure an accurate
isolation of the $N{\frac{1}{2}}^{-}$ state.   
\end{abstract}


\end{frontmatter}

\section{Introduction}

Lattice QCD is very successful in computing many properties of hadrons
from first principles. In particular, the ground state of the hadron
spectrum is a well understood problem~\cite{Durr:2008zz}. However, the
excited states still prove a significant challenge.
One of the long-standing puzzles in hadron spectroscopy has been the
low mass of the first positive parity excitation of the nucleon, known
as the Roper resonance, $N{\frac{1}{2}}^{+}$(1440
MeV)${\rm{P}}_{11}$, compared to the lowest-lying negative parity
partner, $N{\frac{1}{2}}^{-}$ (1535
MeV)${\rm{S}}_{11}$. This phenomenon cannot be observed in
constituent or valence quark models where the lowest-lying odd parity
state naturally occurs below the $N={\frac{1}{2}}^{+}$ state.
Similar difficulties in the level orderings also appear for the
$J^{P}={\frac{3}{2}}^{+} \Delta^{\ast}(1600)$ and ${\frac{1}{2}}^{+}
\Sigma^{\ast} (1690)$ resonances.

 There has been extensive research  focussing on the issue
 of the level  ordering
 problem using the lattice QCD approach
 ~\cite{Sasaki:2001nf,Richards:2001bx,Lee:2002gn,Sasaki:2003xc,Melnitchouk:2002eg,Brommel:2003jm,Mathur:2003zf,Sasaki:2005ap,Guadagnoli:2004wm,Basak:2007kj,Bulava:2009jb,Bulava:2010yg,Engel:2010my}.                                   
One of the state-of-the-art approaches that has been used
extensively in hadron spectroscopy is the `variational method' ~\cite{Michael:1985ne,Luscher:1990ck}, which is
based on a correlation matrix analysis. In the past the
isolation of the Roper resonance was elusive with this method. However, in Refs.~\cite{Mahbub:2009aa,Mahbub:2010jz} a low-lying Roper
state has been identified using a correlation matrix construction with
smeared-smeared correlators. Our work there
motivates us to investigate the long-standing level ordering problem using
the same techniques on the same lattice.

In contrast to the positive parity ground state of the nucleon,
$N{\frac{1}{2}}^{+}$, which has a large plateau over Euclidean-time,
the correlation functions for the negative
parity ground state, $N{\frac{1}{2}}^{-}$, are short-lived
 giving shorter plateaus at earlier Euclidean times.
Therefore, the standard analysis to extract the
$N{\frac{1}{2}}^{-}$ ground  state from small Euclidean-times may
provide a  mixture of ground and excited
 states. On the other hand, the  variational method accounts for the
 presence of  excited  states in the correlation
 functions via correlation matrices. The masses of the energy states are
 then obtained by projecting the correlation matrix to
 eigenstates~\cite{Mahbub:2009aa} providing a     
 robust approach for extracting the energy states.
 In addition, considering several  bases in  constructing different
 correlation matrices provides a substantial verification of the analysis
 technique, in allowing  the consistency of the energy states over the
 different basis  and the reliability of the extracted eigenstates
 energies to be explored.

In this paper, we use the same approach as
that of Ref.~\cite{Mahbub:2009aa} to isolate the negative parity states of
the nucleon. In particular, we focus on the negative parity state to
explore the level ordering problem. \\ 
Various sweeps of gauge invariant Gaussian smearing
~\cite{Gusken:1989qx} are used to construct 
a smeared-smeared correlation function basis to form correlation matrices.

This paper is arranged as follows:
Section~\ref{sec:variational_method} contains a brief description
of the variational method. Lattice details are in
Section~\ref{sec:lattice_details}. Results are
discussed in Section~\ref{sec:results} and
conclusions are presented in Section~\ref{sec:conclusions}.

\section{Variational Method}
 \label{sec:variational_method}

The two point correlation function matrix for $\vec{p} =0$ can be written as
\begin{align}
G_{ij}^{\pm}(t) &= \sum_{\vec x}{\rm Tr}_{\rm sp}\{ \Gamma_{\pm}\langle\Omega\vert\chi_{i}(x)\bar\chi_{j}(0)\vert\Omega\rangle\}, \\
          &=\sum_{\alpha}\lambda_{i}^{\alpha}\bar\lambda_{j}^{\alpha}e^{-m_{\alpha}t},
\end{align}
where, Dirac indices are implicit. Here, $\lambda_{i}^{\alpha}$ and
$\bar\lambda_{j}^{\alpha}$ are the couplings of interpolators $\chi_{i}$ and
$\bar\chi_{j}$ at the sink and source respectively and $\alpha$
enumerates the energy eigenstates with mass
$m_{\alpha}$. $\Gamma_{\pm}=\frac{1}{2}(\gamma_{0}\pm 1)$ projects the parity of the eigenstates.\\ 
 Since the only $t$ dependence comes from the exponential term, one
 can seek a linear superposition of interpolators,
 ${\bar\chi}_{j}u_{j}^{\alpha}$, such that,  
\begin{align}
G_{ij}(t_{0}+\triangle t)\, u_{j}^{\alpha} & = e^{-m_{\alpha}\triangle
  t}\, G_{ij}(t_{0})\, u_{j}^{\alpha},
\end{align}  
for sufficiently large $t_{0}$ and $t_{0}+\triangle t$. More detail can be found in Refs.~\cite{Melnitchouk:2002eg,Mahbub:2009nr,Blossier:2009kd}. Multiplying the above equation by $[G_{ij}(t_{0})]^{-1}$ from the left leads to an eigenvalue equation,
\begin{align}
[(G(t_{0}))^{-1}G(t_{0}+\triangle t)]_{ij}\, u^{\alpha}_{j} & = c^{\alpha}\, u^{\alpha}_{i},
 \label{eq:right_evalue_eq}
\end{align} 
where $c^{\alpha}=e^{-m_{\alpha}\triangle t}$ is the eigenvalue. Similar to Eq.(\ref{eq:right_evalue_eq}), one can also solve the left eigenvalue equation to recover the $v^{\alpha}$ eigenvector,
\begin{align}
v^{\alpha}_{i}\, [G(t_{0}+\triangle t)(G(t_{0}))^{-1}]_{ij} & = c^{\alpha}v^{\alpha}_{j}.
\label{eq:left_evalue_eq}
\end{align} 
The vectors $u_{j}^{\alpha}$ and $v_{i}^{\alpha}$  diagonalize the correlation matrix at time $t_{0}$ and $t_{0}+\triangle t$ making the projected correlation matrix,
\begin{align}
v_{i}^{\alpha}G_{ij}^{\pm}(t)u_{j}^{\beta} & \propto \delta^{\alpha\beta}.
 \label{projected_cf} 
\end{align} 
The parity projected, eigenstate projected correlator, 
\begin{align}
 G^{\alpha}_{\pm}& \equiv v_{i}^{\alpha}G^{\pm}_{ij}(t)u_{j}^{\alpha} ,
 \label{projected_cf_final}
\end{align}
 is then analyzed using standard techniques to obtain the masses of
 different  states.

\section{Simulation Details}
 \label{sec:lattice_details}
 Our lattice ensemble is the same as that explored in
 Ref.~\cite{Mahbub:2009aa}. It consists of 200 quenched configurations
 with a  lattice
volume of $16^{3}\times 32$. Gauge field configurations are generated
by using the DBW2 gauge action~\cite{Takaishi:1996xj,deForcrand:1999bi} and an
${\cal{O}}(a)$-improved FLIC fermion action~\cite{Zanotti:2001yb} is
used to generate quark propagators. This action has excellent scaling
properties and provides near continuum results at finite lattice
spacing~\cite{Zanotti:2004dr}. The lattice spacing is $a=0.127$ fm,
as determined by the static quark potential, with the scale set using
the Sommer scale, $r_{0}=0.49$ fm~\cite{Sommer:1993ce}. In the
irrelevant operators of the fermion action we apply four sweeps of
stout-link smearing to the gauge links to reduce the coupling with the
high frequency modes of the theory~\cite{Morningstar:2003gk}
providing ${\cal O}(a)$ improvement~\cite{Zanotti:2004dr}. We use
the same method as in Refs.~\cite{Lasscock:2005kx,Mahbub:2009nr} to
determine  fixed boundary effects, and the effects are significant
only after  time slice 25 in the present analysis. Various
sweeps of gauge invariant Gaussian smearing~\cite{Gusken:1989qx} (1,
3, 7, 12, 16, 26, 35, 48 sweeps) corresponding to rms 
  radii, in lattice units, of 0.6897, 1.0459, 1.5831, 2.0639, 2.3792,
  3.0284, 3.5237, 4.1868,  are applied at the source ($t=4$)
and at the sink. This is to ensure a large range of overlaps of the
interpolators with the lower-lying states. 
The analysis is performed on eight different quark masses
corresponding to pion masses of
$m_{\pi}=\{0.797,0.729,0.641,0.541,0.430,0.380,0.327,0.295\}$ GeV.
The error analysis is performed using the jackknife method,
with the ${\chi^{2}}/{\rm{dof}}$ obtained via a covariance matrix
analysis method. Our fitting method is discussed extensively in Ref.~\cite{Mahbub:2009nr}.\\
The nucleon interpolators we consider are,      
\begin{align}
\chi_1(x) &= \epsilon^{abc}(u^{Ta}(x)\, C{\gamma_5}\, d^b(x))\,u^{c}(x), \\
\chi_2(x) &= \epsilon^{abc}(u^{Ta}(x)Cd^b(x)){\gamma_5}\, u^{c}(x). 
\label{eqn:chi_1_2_interpolator}
\end{align}

We use the Dirac representation of the gamma matrices in our analysis.

\section{Results}
 \label{sec:results}

\begin{figure*}[!t] 
  \begin{center}
   $\begin{array}{c@{\hspace{0.15cm}}c}  
 \includegraphics [height=0.45\textwidth,angle=90]{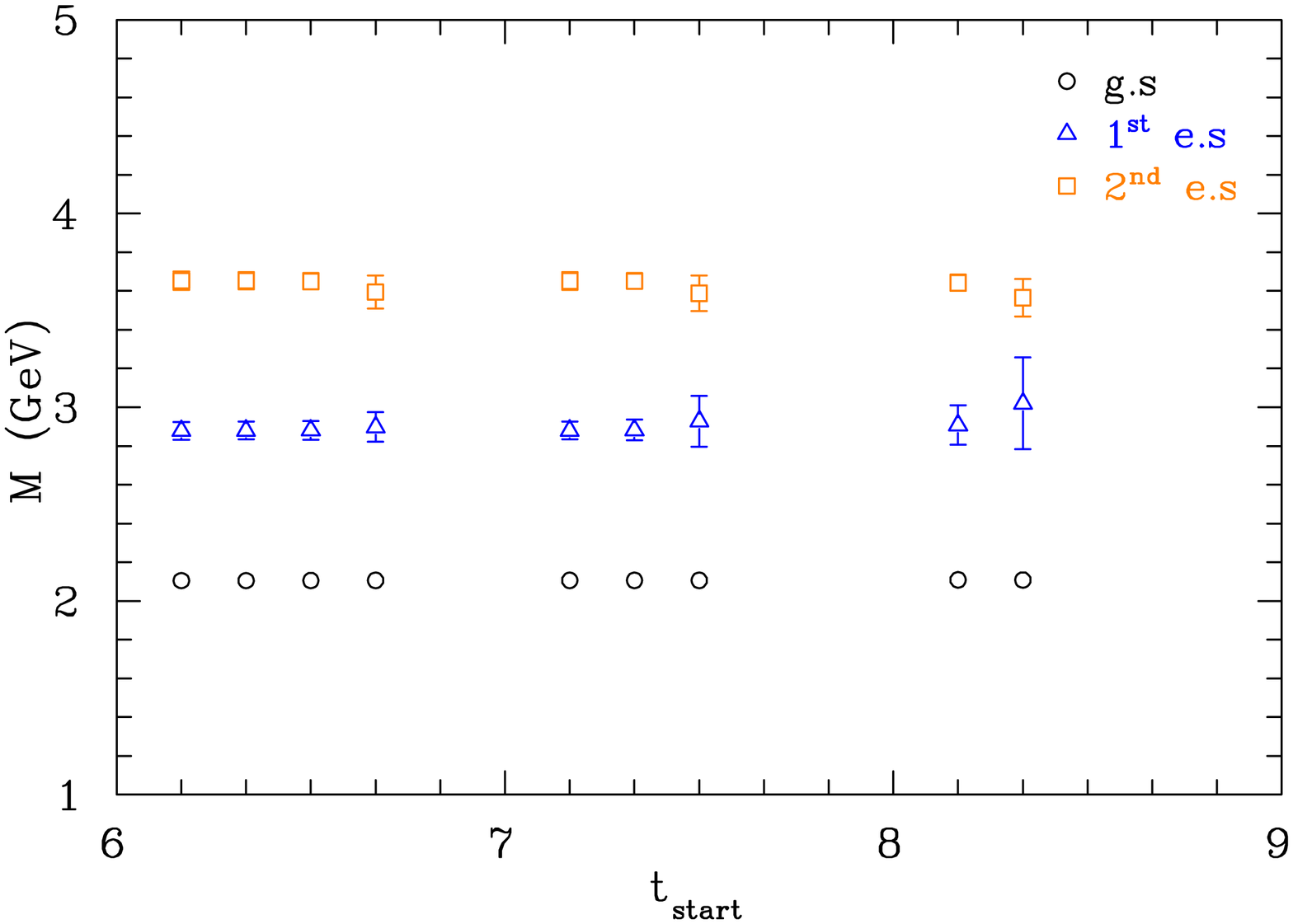} &
 \includegraphics [height=0.45\textwidth,angle=90]{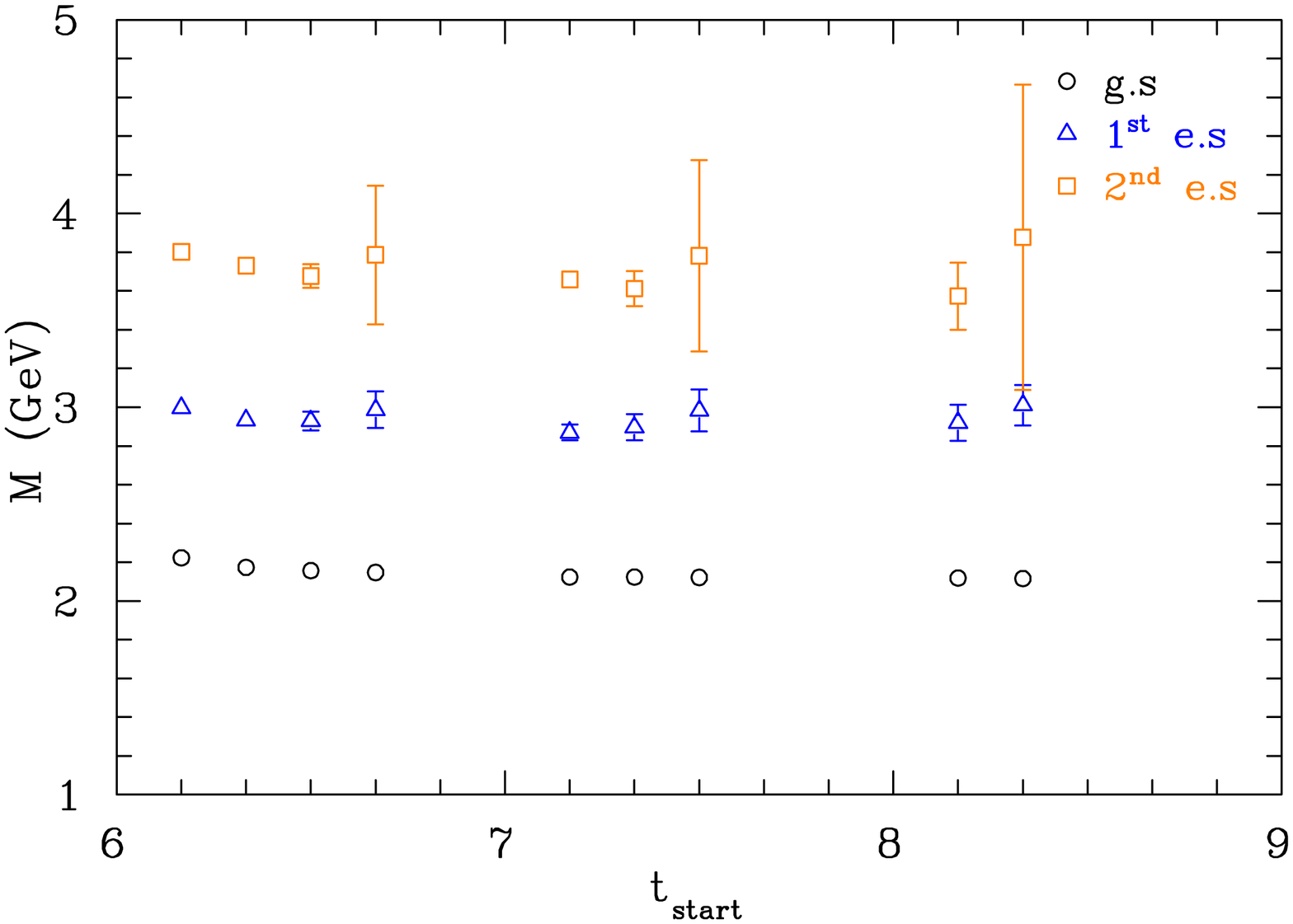}    
      \vspace{0.10cm} 
    \end{array}$

 $\begin{array}{c@{\hspace{0.15cm}}c}  
 \includegraphics [height=0.45\textwidth,angle=90]{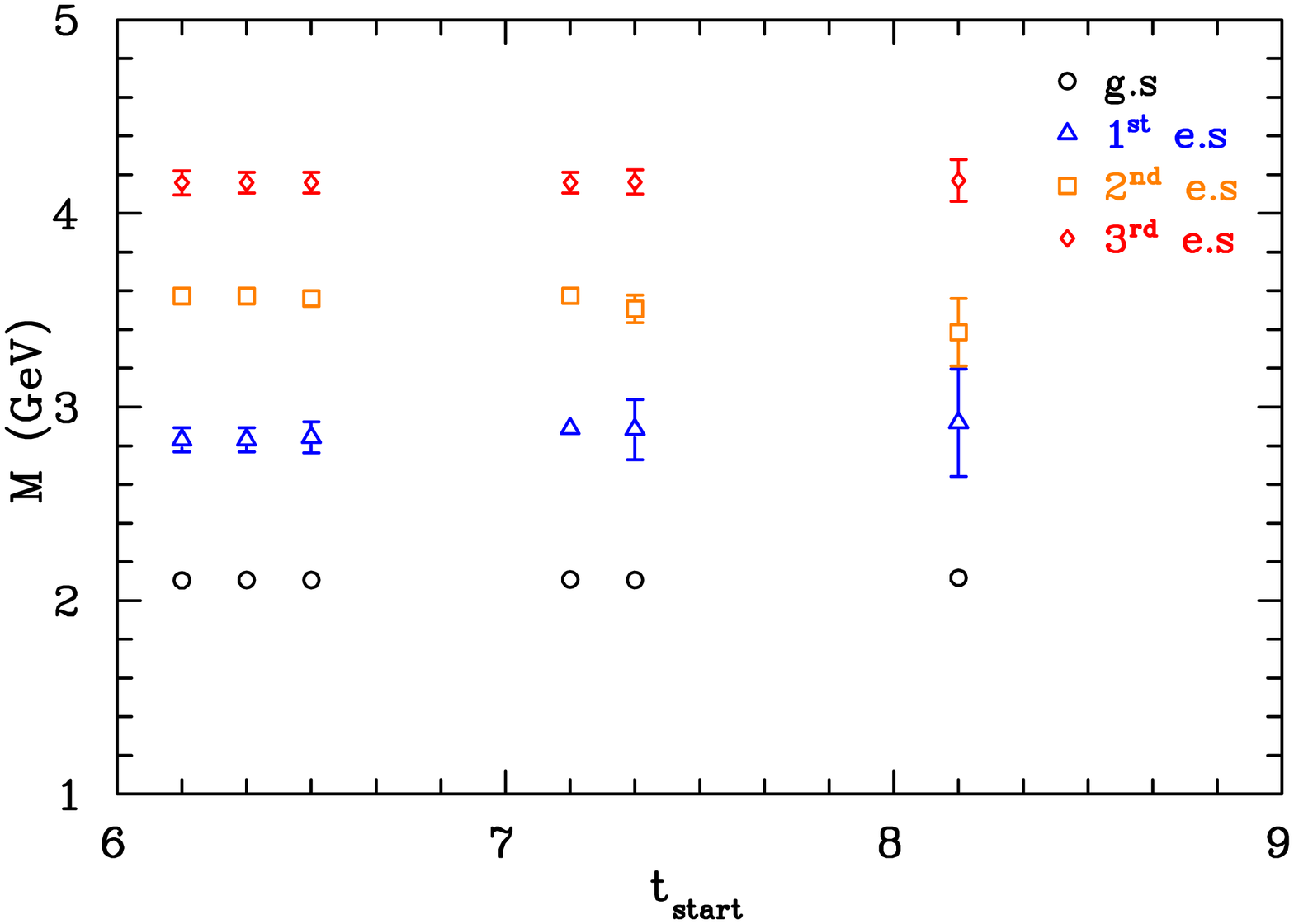} &
 \includegraphics [height=0.45\textwidth,angle=90]{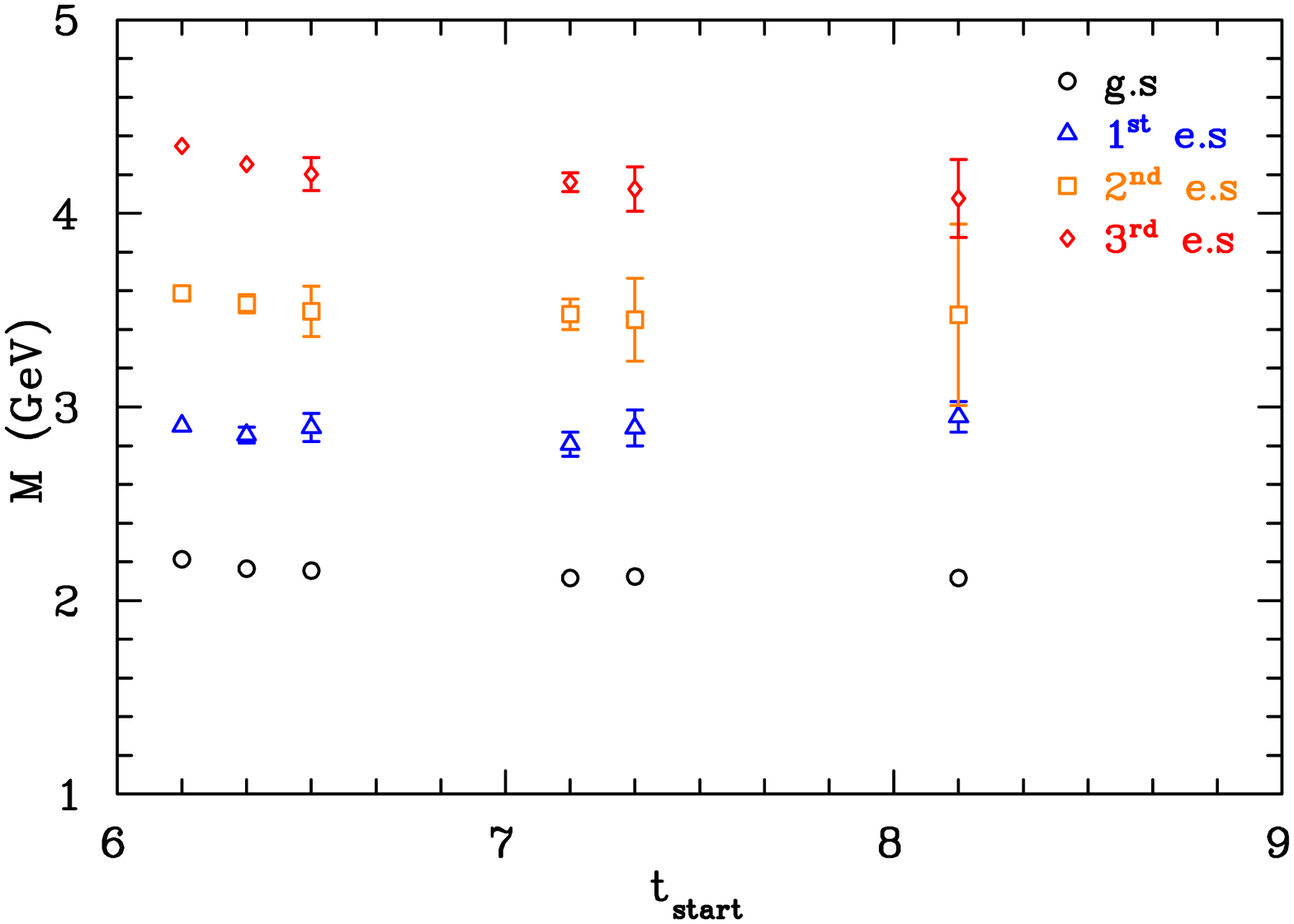}    
      \vspace{0.10cm} 
    \end{array}$

\vspace{-0.25cm}
    \caption{(colour online). Masses of the nucleon, $N{\frac
          {1}{2}}^{-}$ states,  from the projected correlation
      functions  (left) and from the
        eigenvalues (right), for the $3\times 3$ (top row) and
        $4\times 4$ (bottom row) correlation matrices of
        $\chi_{1}\bar\chi_{1}$, for the pion mass of 797 MeV. The
        $3\times 3$  results correspond to
        the $1^{\rm{st}}$ combination (7,16,26 sweeps) of $3\times 3$
        matrices, while $4\times 4$ results correspond to the
        $2^{\rm nd}$ combination (3,12,26,35 sweeps) of $4\times 4$
        matrices. Each set of $N{\frac{1}{2}}^{-}$ ground (g.s) and
        excited (e.s) states masses correspond to the
       diagonalization of the correlation matrix for each set of
       variational  parameters $t_{\rm start}$ (shown in major tick
       marks) and $\triangle t$ (shown in minor tick marks). Here,
       $t_{\rm start}\equiv t_{0}$ as shown in
       Eqs.~(\ref{eq:right_evalue_eq}) and~(\ref{eq:left_evalue_eq}).}  
   \label{fig:mass_and_eig_3x3_4x4_x1x1_Q1}
  \end{center}
\end{figure*}

We consider several $3\times 3$, $4\times 4$, $6\times 6$ and
$8\times8$ correlation matrices. Each matrix is constructed with
different  sets of
correlation functions, each set element corresponding to a different
numbers of sweeps of gauge invariant Gaussian smearing at the 
source and sink of the $\chi_{1}\bar\chi_{1}$, $\chi_{2}\bar\chi_{2}$ and
$\chi_{1}\chi_{2}$ correlators~\cite{Mahbub:2010jz}. This provides a
large  basis of operators with a variety of overlaps among energy
states. 

We consider five smearing combinations (bases) \{1=(7,16,26), 2=(7,16,35),
3=(12,16,26), 4=(12,26,35), 5=(16,26,35)\} for $3\times 3$ correlation
function matrices and four combinations \{1=(1,12,26,48), 2=(3,12,26,35),
3=(3,12,26,48), 4=(12,16,26,35)\} for $4\times 4$
matrices, of $\chi_{1}\bar\chi_{1}$ correlation
  functions. In the latter case  these four
combinations are found optimal for the reliable extraction of the
low-lying energy states shown in Ref.~\cite{Mahbub:2009aa}. Including the
$\chi_{2}$ interpolator, which vanishes in the
non-relativistic limit ~\cite{Leinweber:1990dv,Leinweber:1994nm}, in
correlation matrix analysis provides extra challenges.
Nonetheless, we consider this interpolator for the reliable
extraction of the negative parity ground state mass. The same bases, as
discussed above for the $\chi_{1}\bar\chi_{1}$ analysis, are also
considered for the $3\times 3$ and $4\times 4$ correlation matrices of
$\chi_{2}\bar\chi_{2}$ correlation functions. We also consider four
smearing  combinations \{1=(3,12,26), 2=(3,16,48), 3=(7,16,35),
4=(12,16,26)\} of $6\times 6$ and four combinations \{1=(3,12,26,48),
2=(7,12,26,35),  3=(7,16,26,35), 4=(7,16,35,48)\} for $8\times 8$
matrices of  $\chi_{1}\chi_{2}$ correlation functions.

 \begin{figure*}[!t]
  \begin{center}
   $\begin{array}{c@{\hspace{0.80cm}}c}  
 \includegraphics [height=0.45\textwidth,angle=90]{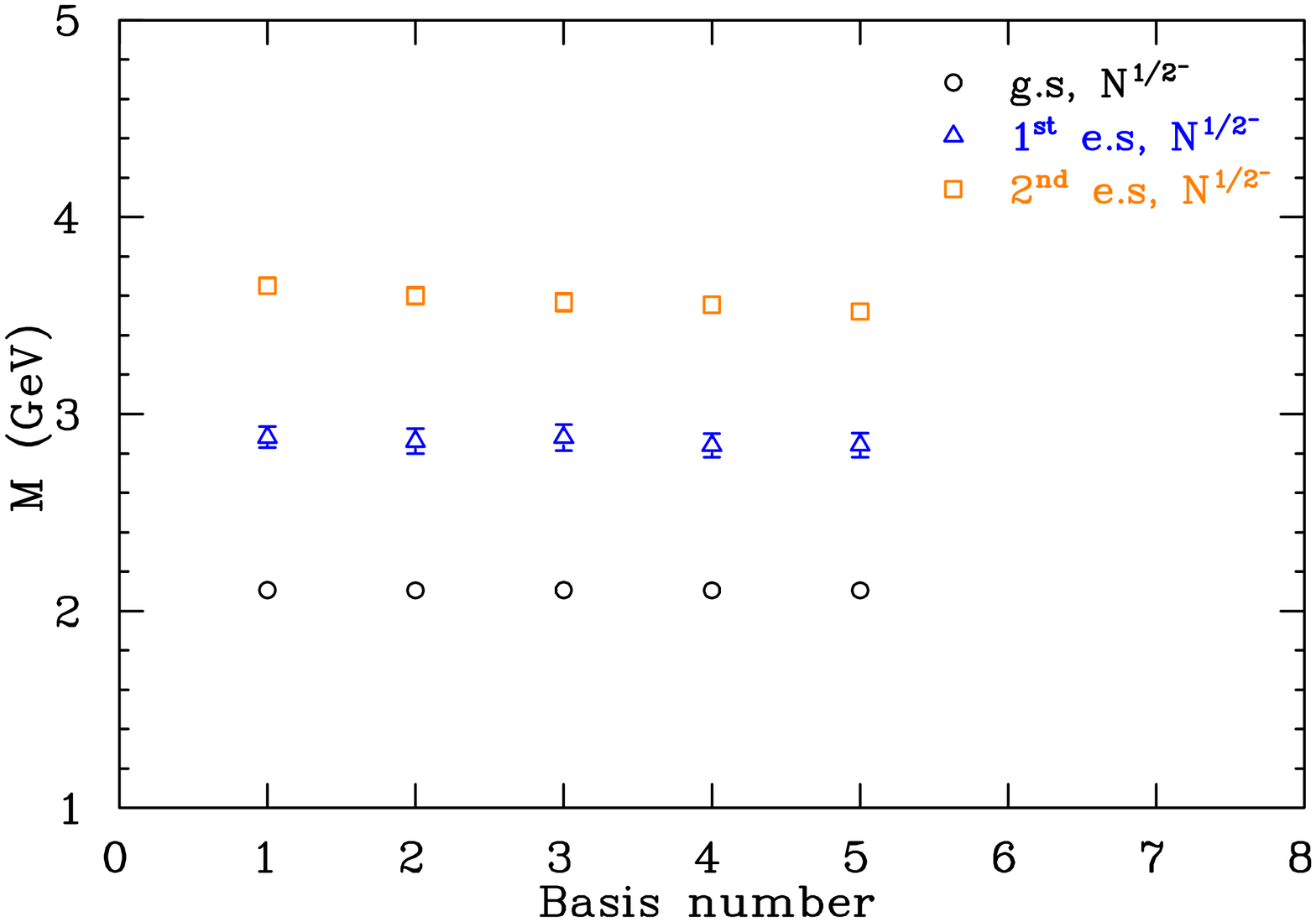} &
 \includegraphics [height=0.45\textwidth,angle=90]{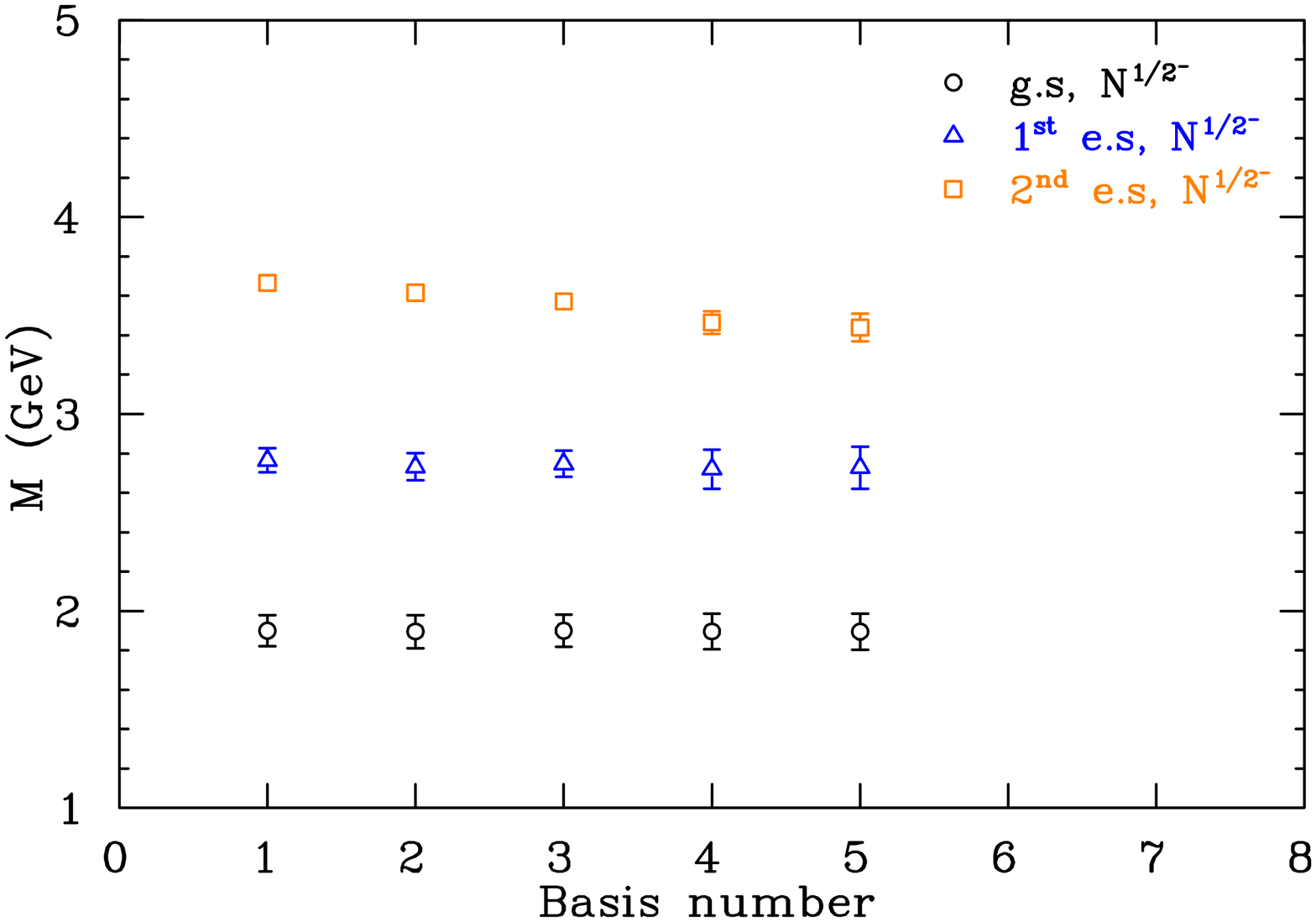}        \end{array}$
    \caption{(color online). Masses of the nucleon, $N{\frac
           {1}{2}}^{-}$ - states, from projected correlation
      functions as shown in Eq.~(\ref{projected_cf_final})
       for the pion mass of 797 MeV (left) and 380 MeV (right). Numbers in
       the horizontal scale correspond to each combination of smeared
       $3\times 3$ correlation matrices of $\chi_{1}\bar\chi_{1}$. For instance,
       1 and 2 correspond to the combinations of (7,16,26) and (7,16,35)
       sweeps, respectively and so on, as
       discussed in the text following
       Eq.~(\ref{eqn:chi_1_2_interpolator}).} 
   \label{fig:m_all_comb_3x3_Q1-6}
  \end{center}
\end{figure*}

 \begin{figure*}[!t]
  \begin{center}
   $\begin{array}{c@{\hspace{0.80cm}}c}  
 \includegraphics [height=0.45\textwidth,angle=90]{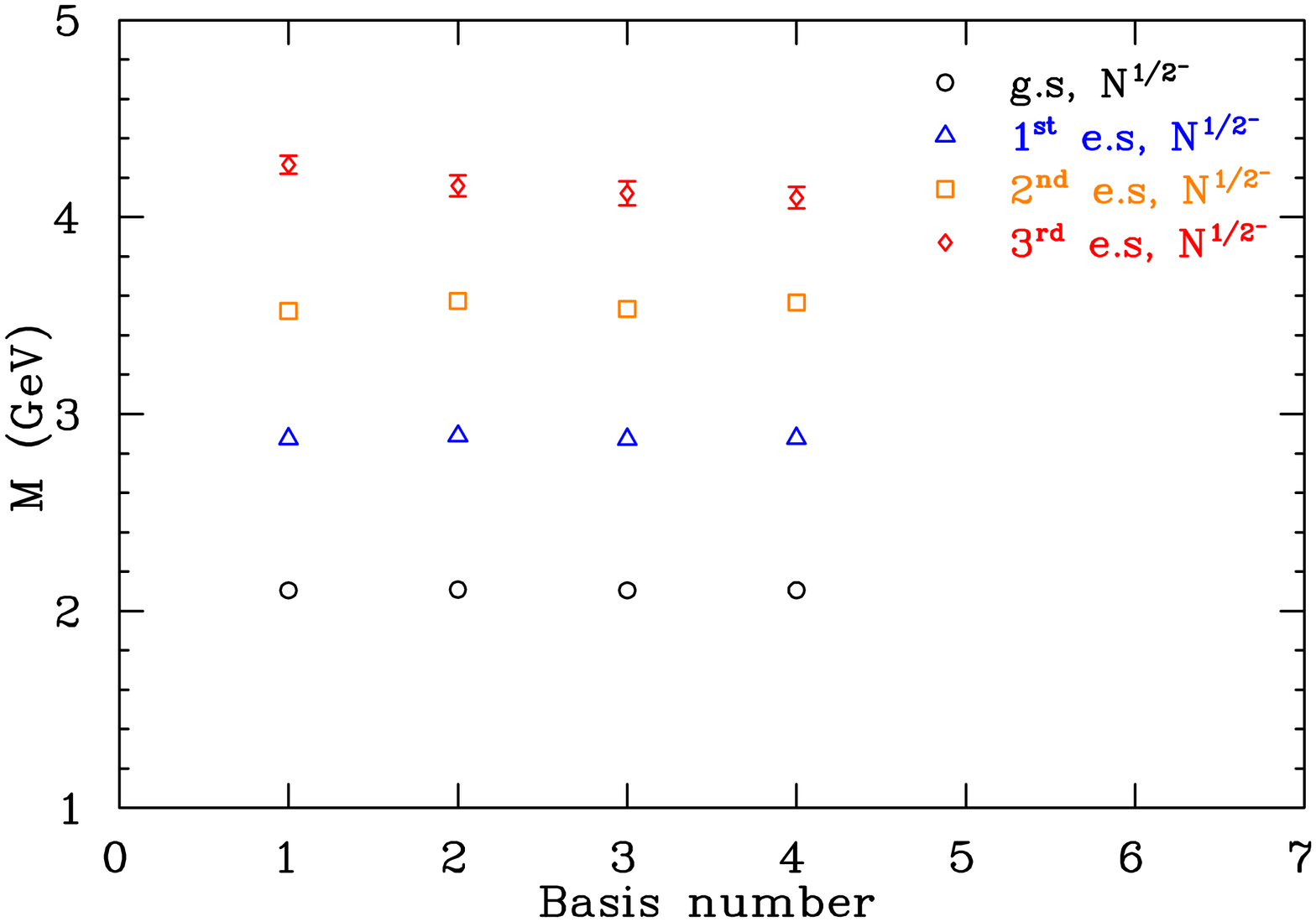} &
 \includegraphics [height=0.45\textwidth,angle=90]{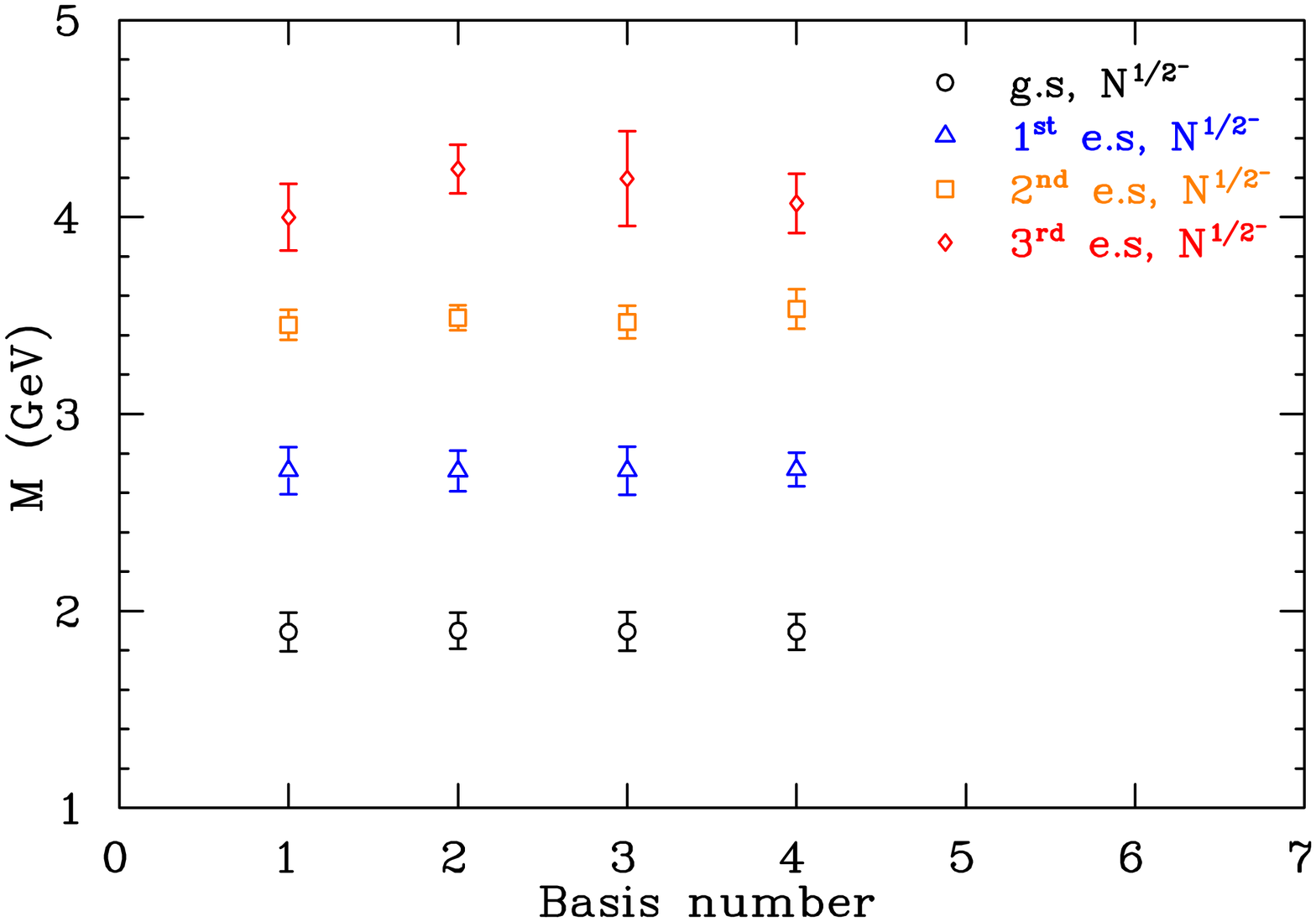}        \end{array}$
    \caption{(color online). As in Fig.~\ref{fig:m_all_comb_3x3_Q1-6},
      but for the $4\times 4$ correlation matrices of $\chi_{1}\bar\chi_{1}$.}       
   \label{fig:m_all_comb_4x4_Q1-6}
  \end{center}
\end{figure*}

In Fig.~\ref{fig:mass_and_eig_3x3_4x4_x1x1_Q1}, masses from the
projected correlation functions and from the eigenvalues are shown for
the $3\times 3$ and $4\times 4$ correlation matrices. We refer to the
lowest lying state as the ground state in the negative parity sector. As in
Refs.~\cite{Mahbub:2009nr,Mahbub:2009aa}, masses from
the projected correlation functions for the low-lying
states are very consistent over the variational parameters of
 $t_{\rm start}$  and $\triangle t$, in particular, the negative
parity ground state is robust. However, a deteriorating signal to
noise is evident for larger $t_{\rm start}$ and
$\triangle t$ values, particularly for the excited states. In
contrast, the mass  from the eigenvalue analysis shows significant
dependence on the variational parameters. Therefore, exposing a mass
from the projected correlation functions is again proved to be more
reliable than from the eigenvalues~\cite{Mahbub:2009aa}.

From a series of $t_{\rm start}$
 and $\triangle t$, a single mass is 
selected for one set of $t_{\rm start}$ and $\triangle t$ by the
selection criteria discussed in Ref.~\cite{Mahbub:2009nr}, where we
prefer larger value of $t_{\rm start}+\triangle
t$~\cite{Blossier:2009kd}.  In cases where a
larger $t_{\rm start}+\triangle t$ provides a poor signal-to-noise
ratio, for example ($t_{\rm start},\triangle t$)=(7,3) (top left
graph of Fig.~\ref{fig:mass_and_eig_3x3_4x4_x1x1_Q1}), we prefer a little lower $t_{\rm start}+\triangle t$ value,
for example ($t_{\rm start},\triangle t$)=(7,2), and we follow this
procedure for each quark mass, as discussed in Ref.~\cite{Mahbub:2010jz}.

In Fig.~\ref{fig:m_all_comb_3x3_Q1-6}, masses extracted from
all the combinations of $3\times 3$ matrices (from $1^{\rm st}$ to
$5^{\rm th}$) are shown for the pion mass of 797 and 380 MeV. Here the negative
parity ground and the first excited states are very consistent for all
the $3\times 3$ bases. The second excited state shows some smearing
dependence ~\cite{Mahbub:2009nr} which is also observed in
Ref.~\cite{Mahbub:2009aa} as this highest excited state must
accommodate all remaining spectral strength.

In Fig.~\ref{fig:m_all_comb_4x4_Q1-6}, masses extracted from
all the combinations of $4\times 4$ matrices of $\chi_{1}\bar\chi_{1}$
are  shown. In
Ref.~\cite{Mahbub:2009aa}, the three low-lying states were consistent
over these four bases for the positive parity excited states and 
similarly consistent results are also observed in this negative parity
correlation matrix analysis, again illustrating the robustness of the
method~\cite{Mahbub:2009aa}.  A smearing dependency for the
highest excited state is also noticed in this $4\times 4$ analysis,
which is to be expected.

\begin{figure}[!ht]
  \begin{center}
 \includegraphics[height=0.45\textwidth,angle=90]{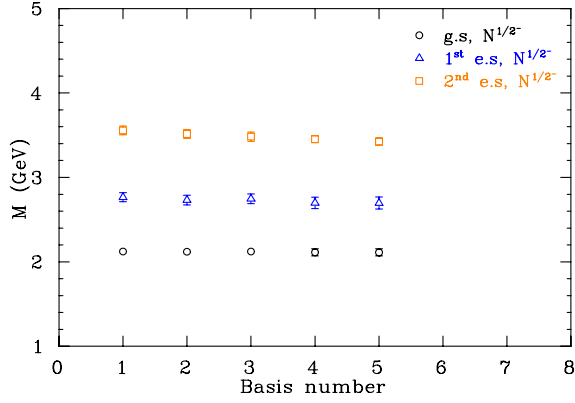} 
    \caption{(color online). Masses of the nucleon, $N{\frac
           {1}{2}}^{-}$ - states, from projected correlation
      functions as shown in Eq.~(\ref{projected_cf_final})
       for the pion mass of 797 MeV. Numbers in
       the horizontal scale correspond to each combination of smeared
       $3\times 3$ correlation matrices of $\chi_{2}\bar\chi_{2}$. For instance,
       1 and 2 correspond to the combinations of (7,16,26) and (7,16,35)
       sweeps, respectively and so on.} 
   \label{fig:m_negP_sosi_3x3_x2x2_Q1}
  \end{center}
\end{figure}

\begin{figure}[!ht]
  \begin{center}
 \includegraphics[height=0.45\textwidth,angle=90]{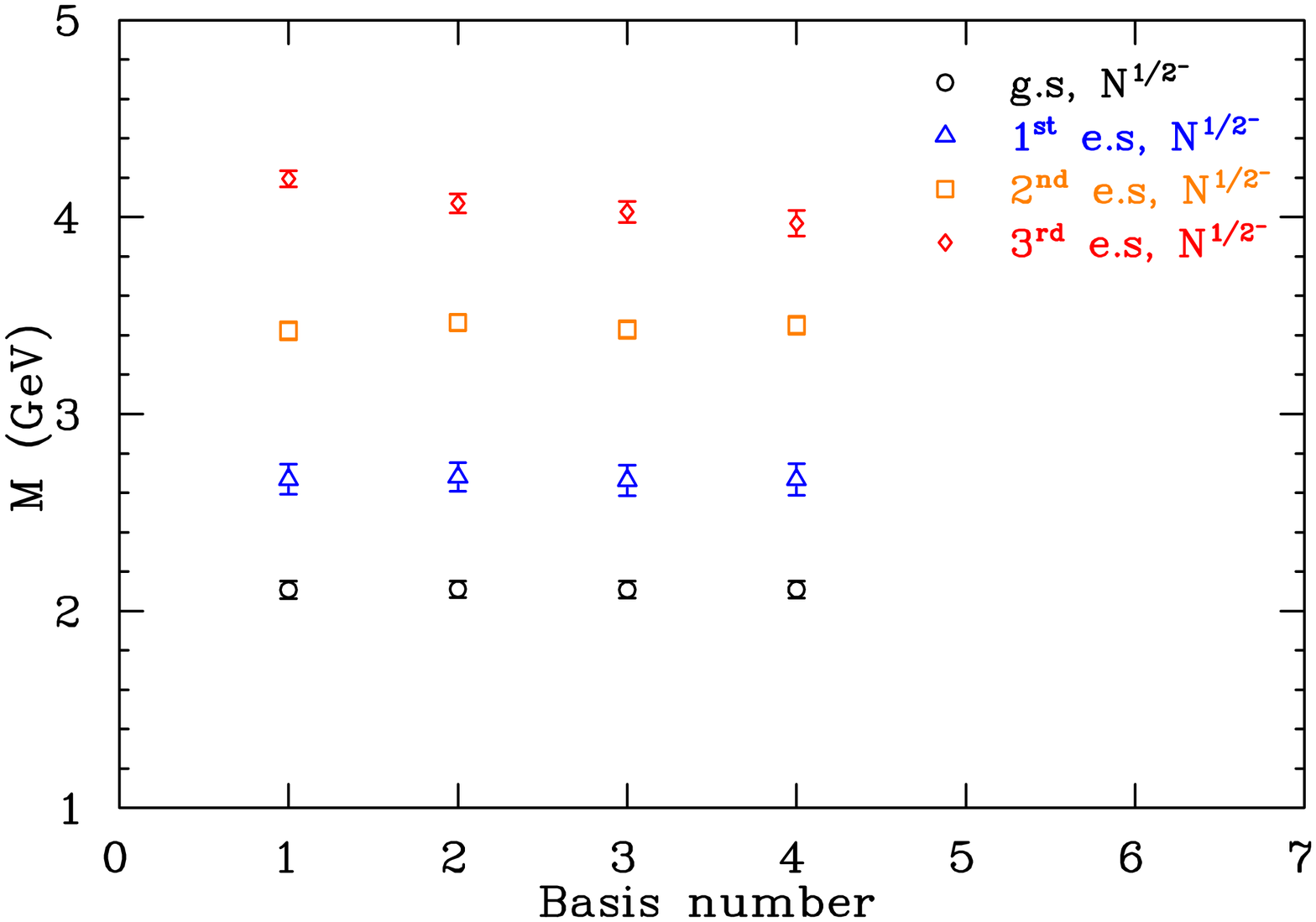} 
    \caption{(color online). Masses of the nucleon, $N{\frac
           {1}{2}}^{-}$ - states, from projected correlation
      functions as shown in Eq.~(\ref{projected_cf_final})
       for the pion mass of 797 MeV. Numbers in
       the horizontal scale correspond to each combination of smeared
       $4\times 4$ correlation matrices of $\chi_{2}\bar\chi_{2}$. For instance,
       1 and 2 correspond to the combinations of (1,12,26,48) and (3,12,26,35)
       sweeps, respectively and so on.} 
   \label{fig:m_negP_sosi_4x4_x2x2_Q1}
  \end{center}
\end{figure}

 It is noted that although the $3\times 3$
correlation matrix analyses are  successful for all the eight quark
masses, the $4\times 4$ correlation matrix analyses
are successful only for the six heavier quarks. In this case, the
signal-to-noise
ratio for the lighter quarks deteriorates more rapidly than
the heavier quarks. As a result the variational analysis for $4\times 4$
correlation matrices for the lighter two quarks is unsuccessful.

We now discuss the $3\times 3$ and $4\times 4$ correlation matrices of
$\chi_{2}\bar\chi_{2}$. As the $\chi_{2}$ interpolator is very
challenging to perform a simulation with and the signal-to-noise ratio
of the $\chi_{2}\bar\chi_{2}$ correlator deteriorates rapidly,
the diagonalization
of $3\times 3$ (Fig.~\ref{fig:m_negP_sosi_3x3_x2x2_Q1}) and $4\times
4$ (Fig.~\ref{fig:m_negP_sosi_4x4_x2x2_Q1}) matrices are only successful
for the four heavier quark masses. The energy states revealed by the
$\chi_{2}$ spin-flavour analysis are also very consistent as with the
$\chi_{1}\bar\chi_{1}$ analysis. In particular, the
$N{\frac{1}{2}}^{-}$ ground state is again robust.

\begin{figure}[!ht]
  \begin{center}
 \includegraphics[height=0.45\textwidth,angle=90]{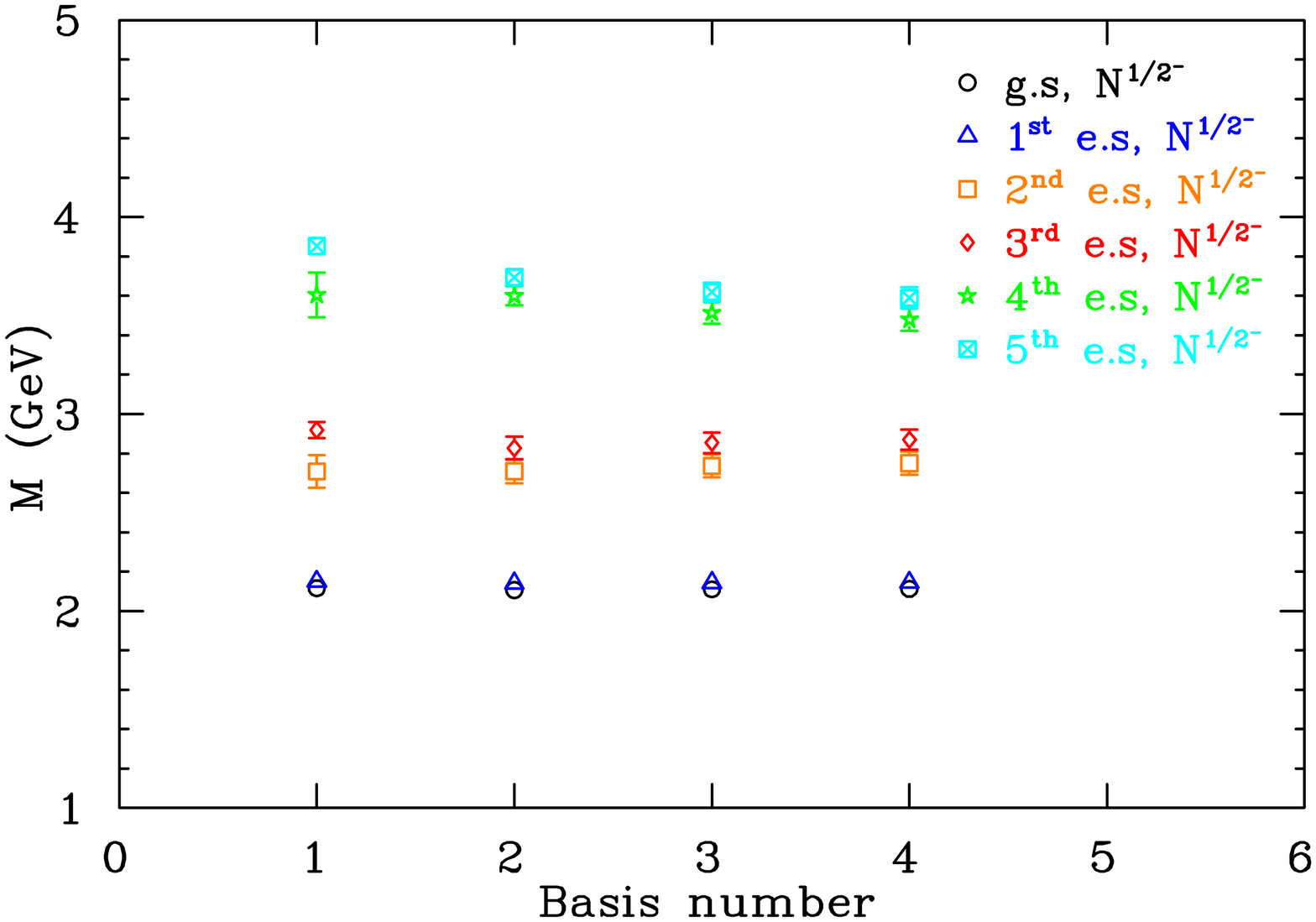} 
    \caption{(color online). Masses of the nucleon, $N{\frac
           {1}{2}}^{-}$ - states, from projected correlation
      functions as shown in Eq.~(\ref{projected_cf_final})
       for the pion mass of 797 MeV. Numbers in
       the horizontal scale correspond to each combination of smeared
       $6\times 6$ correlation matrices of $\chi_{1}\chi_{2}$. For instance,
       1 and 2 correspond to the combinations of (3,12,26) and (3,16,48)
       sweeps, respectively and so on.} 
   \label{fig:m_negP_sosi_6x6_x1x2_Q1}
  \end{center}
\end{figure}

\begin{figure}[!ht]
  \begin{center}
 \includegraphics[height=0.45\textwidth,angle=90]{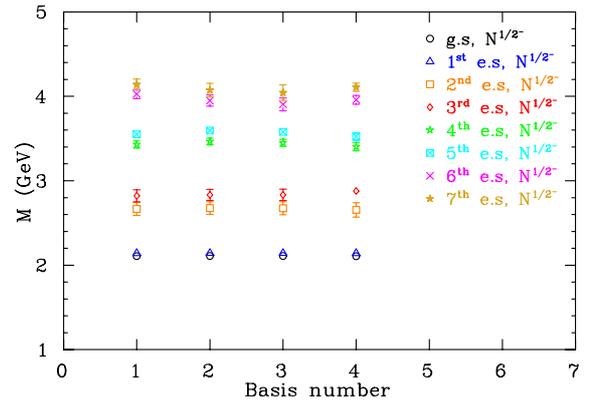} 
    \caption{(color online). As in
      Fig.~\ref{fig:m_negP_sosi_6x6_x1x2_Q1}, but for the $8\times 8$
      correlation matrices of $\chi_{1}\chi_{2}$.} 
   \label{fig:m_negP_sosi_8x8_x1x2_Q1}
  \end{center}
\end{figure}

 \begin{figure*}[!t]
  \begin{center}
   $\begin{array}{c@{\hspace{0.6cm}}c}  
 \includegraphics[height=0.45\textwidth,angle=90]{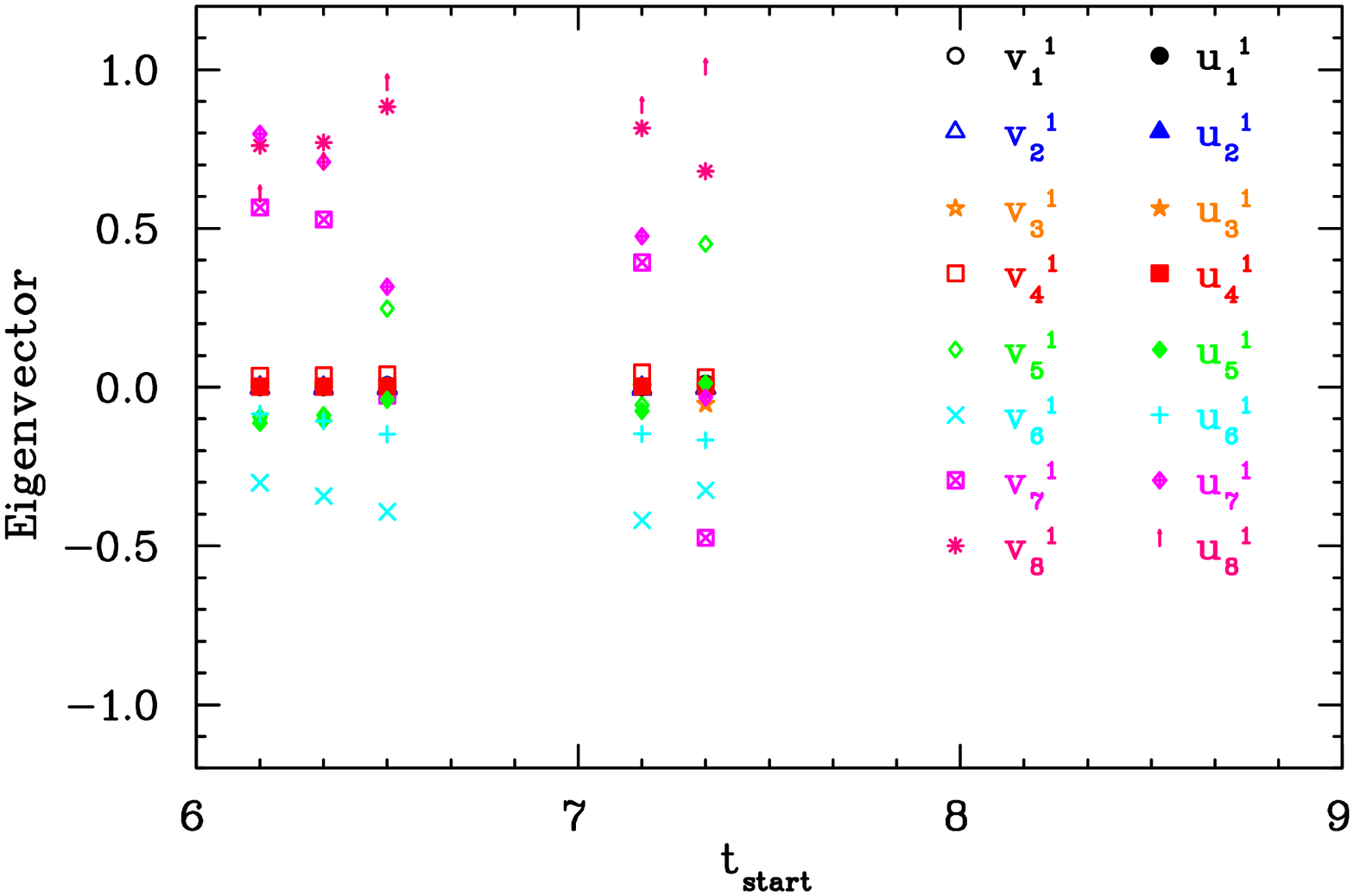}&
 \includegraphics[height=0.45\textwidth,angle=90]{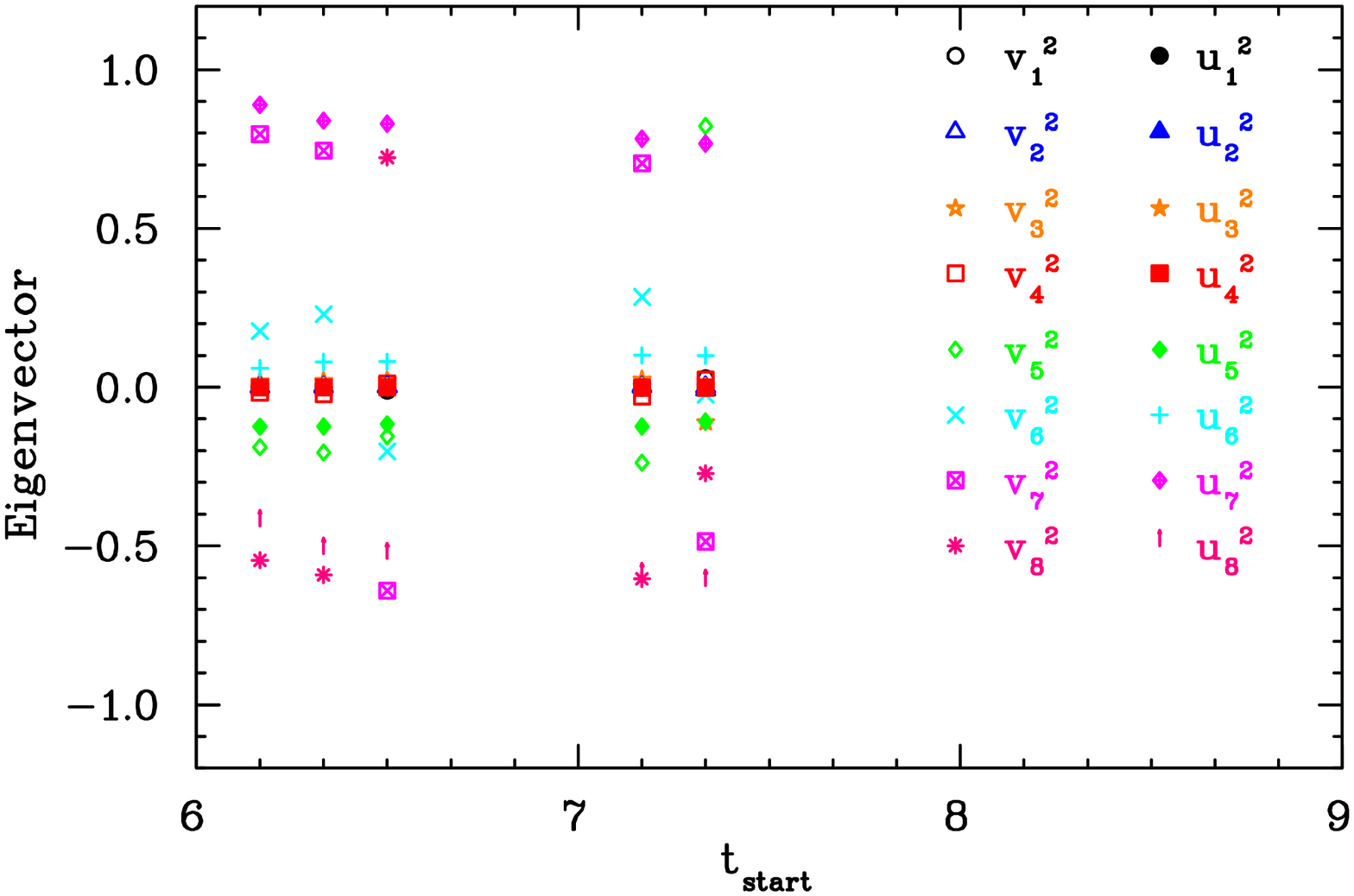}
    \end{array}$
   $\begin{array}{c@{\hspace{0.6cm}}c}  
 \includegraphics[height=0.45\textwidth,angle=90]{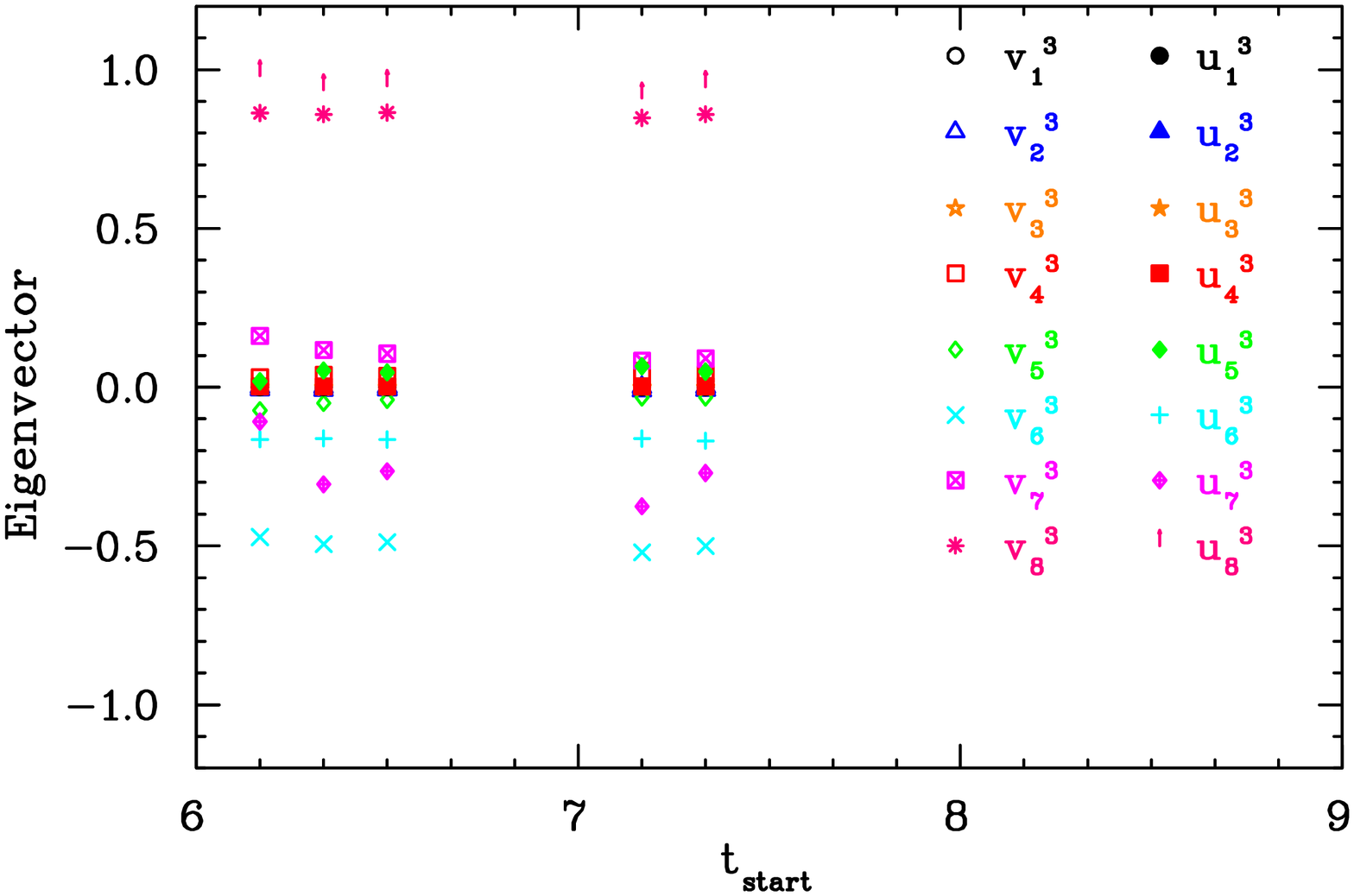}&
 \includegraphics[height=0.45\textwidth,angle=90]{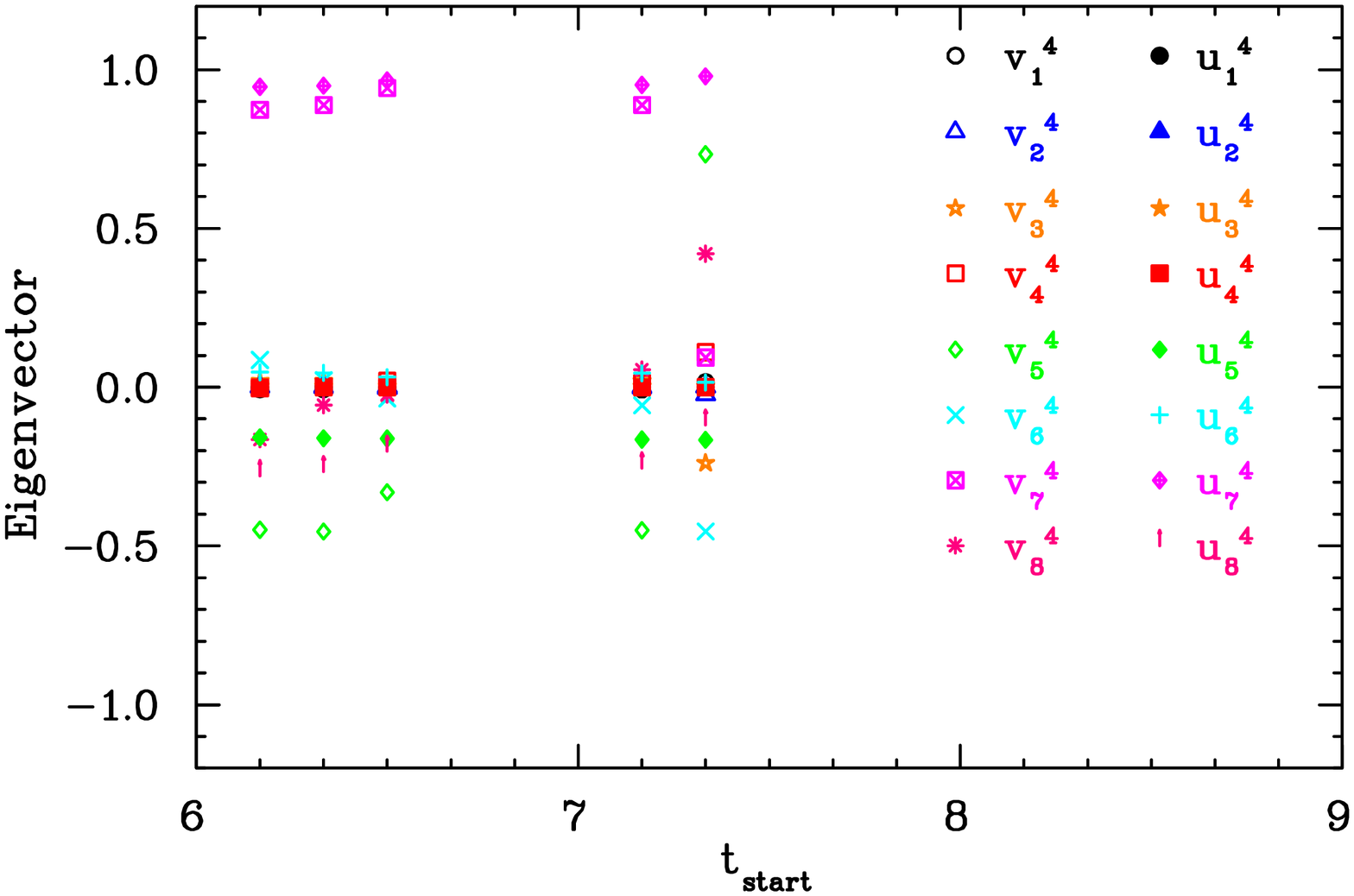}
    \end{array}$
    \caption{(color online). Eigenvector values, $v_{i}^{\alpha}$ and
      $u_{i}^{\alpha}$, as shown in Eq.~(\ref{projected_cf_final}),
      for the first (lowest) state (top left), second (top
      right), third (bottom left) and fourth (bottom right) energy
      states. The figure
      corresponds to the pion mass of 797 MeV and for the $2^{\rm nd}$
      combination (7,12,26,35 sweeps) of $8\times 8$ correlation
      matrix of $\chi_{1}\chi_{2}$. In the legend, superscripts
      stand for the energy state and subscripts represent eigenvector
      contributions of interpolators to make a state of interest,
      e.g.~$v^{1}_{1}$,$v^{1}_{3}$,$v^{1}_{5}$,$v^{1}_{7}$  and
      $v^{1}_{2}$,$v^{1}_{4}$,$v^{1}_{6}$,$v^{1}_{8}$ correspond to
      the contributions from $\chi_{1}$ and
      $\chi_{2}$  operator for the ground state respectively,
      with larger subscript values corresponding to larger smearings.
      To be specific, component $v^{\alpha}_{1}$ and $v^{\alpha}_{2}$
      correspond to the 
      low  smearing sweep count of 7, $v^{\alpha}_{3}$ and
      $v^{\alpha}_{4}$ for the sweep count of 12, and so on. Errors are
      suppressed for clarity.} 
   \label{fig:evectors_8x8_x1x2_st1-2_Q1}
  \end{center}
\end{figure*}

We now consider the $6\times 6$ and $8\times 8$ analyses involving
both $\chi_{1}$ and $\chi_{2}$, where linear combinations of
$\chi_{1}$ and $\chi_{2}$ are utilized to isolate eigenstates.  

The lowest two energy states extracted with the
  $6\times 6$  basis are
nearly degenerate as illustrated in Fig.~\ref{fig:m_negP_sosi_6x6_x1x2_Q1}. An
explanation for these two nearby degenerate states is provided by the quark
model.  In the quark model based on SU(6) symmetry, four decuplet
and two octet states contribute to provide the 56
representation of SU(6). The nucleon and Roper belongs to the ground
and excited 56 representation of SU(6) symmetry respectively. On the
other hand, the odd parity ground (1535 MeV) and (1650 MeV) states
belong to the negative parity, $L=1$, 70 plet of SU(6). As three
spin-$\frac{1}{2}$ quarks may give rise to a total spin of
$s=\frac{1}{2}$ or $\frac{3}{2}$, the $L=1$ state can couple two
different ways to provide a $J=\frac{1}{2}$ state,
i.e. $\vec{1}+\vec{\frac{1}{2}}=\frac{1}{2}$ or
$\vec{1}+\vec{\frac{3}{2}}=\frac{1}{2}$, hence providing two orthogonal spin-$\frac{1}{2}$ states in the $L=1$, 70 plet representation. Both of these
negative-parity states have a width of $\approx$150 MeV.
 It is interesting that
the $\chi_{1}\chi_{2}$ spin flavour combinations successfully isolate
the  nearly degenerate ground states, in accord with the SU(6) quark
model. This approximate degeneracy is repeated throughout our observed spectrum.

In Fig.~\ref{fig:m_negP_sosi_8x8_x1x2_Q1}, similar nearly degenerate
 low-lying states from the $8\times 8$ analysis are evident 
as in the $6\times 6$ case. We note that, only the analysis
  involving $\chi_{2}$ is able to extract  a new second excited state.
  This state is close to the
experimentally identified one star, $N{\frac{1}{2}}^{-}$ (2090 MeV)
$S_{11}$ state which has $\approx$400 MeV of width. Further analysis
is  required  with higher statistics to explore the propagation of these 
states to light quark masses.

\begin{figure*}[!t]
  \begin{center}

 \includegraphics[height=0.80\textwidth,angle=90]{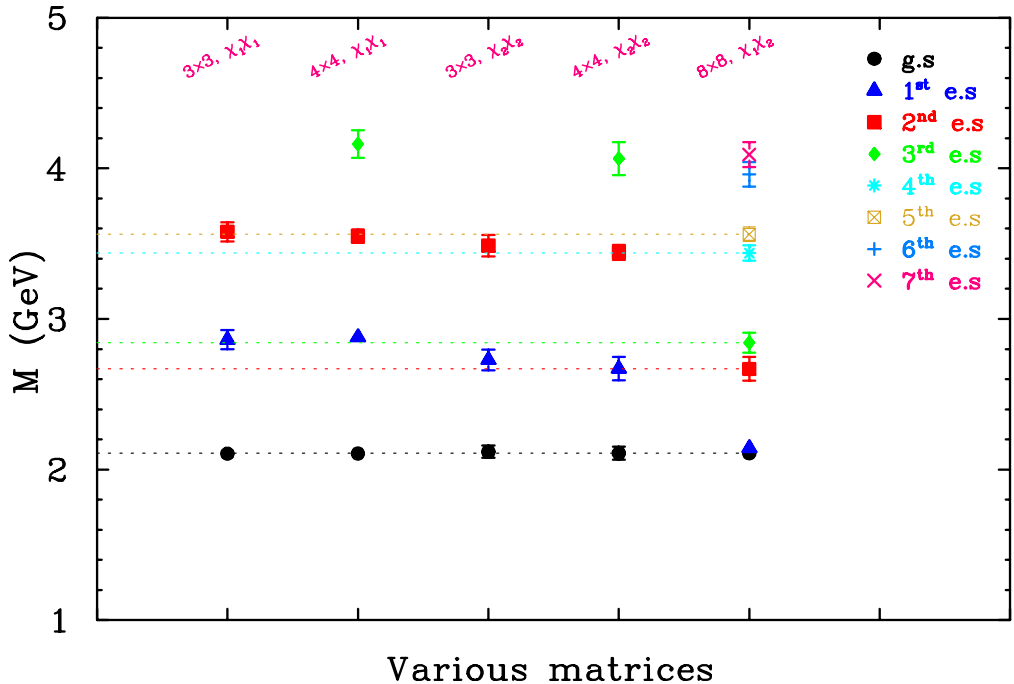} 
    \caption{(color online). Masses of the nucleon, $N{\frac
           {1}{2}}^{-}$ states, for all the various dimensions of
      correlation matrices as labelled on the upper horizontal
      axis. Dotted lines are drawn to aid in illustrating the
      consistency of the results. Figure corresponds to the pion mass
      of 797 MeV.} 
   \label{fig:negP_history_3x3_4x4_6x6_8x8_Q1}
  \end{center}
\end{figure*}

The eigenvectors
  (Fig.~\ref{fig:evectors_8x8_x1x2_st1-2_Q1}) indicate that both
$\chi_{1}$ and $\chi_{2}$ operators have nontrivial contributions in
extracting the two lowest lying states. In particular, the
contributions to the ground state is significantly dominated by
the $\chi_{2}$ interpolator, whereas a larger dominance of the
$\chi_{1}$ interpolator is evident for the first excited state,
revealing a diminished role for the scalar diquark in the negative
parity sector. This observation is in accord with
Ref.~\cite{Engel:2010my}.

 It has already been mentioned that only the
correlation matrix analysis involving $\chi_{2}$ spin-flavour combinations
is able to extract the third energy state. It is evident from the
eigenvector contributions that this energy state is almost a pure
$\chi_{2}$  state (bottom left graph). The fourth energy state is
 dominated by the $\chi_{1}$ operator. It is evident
that larger  smearing, i.e. sweep count of 35, dominates over low
smearing levels in isolating these eigen-energy states. 

 The extracted masses from  all the
bases  are very consistent. In particular,
the  ground state is robust. Therefore, all the bases  are considered
in  performing the following systematic error analysis.

 A systematic error analysis as that of
Ref.~\cite{Mahbub:2009aa} is performed  to calculate
the systematic errors associated with the choice of basis, with
$\sigma_{b}=\sqrt{{\frac{1}{N_{b}-1}}\sum_{i=1}^{N_{b}}(M_{i}-\bar{M})^{2}}$,
where $N_{b}$ is the number of bases.
This analysis is performed over all the bases considered. The masses
are averaged over the bases and errors
(average statistical errors and systematic errors associated with
basis choices) are combined in quadrature,
$\sigma=\sqrt{\bar{\sigma}_{s}^{2}+\sigma_{b}^{2}}$.

At this stage, we examine the consistency of the low-lying energy states
over different dimensions of the correlation
 matrices in Fig.~\ref{fig:negP_history_3x3_4x4_6x6_8x8_Q1}. Whereas a non-trivial mixing of $\chi_{1}$ and $\chi_{2}$
 is required to isolate the two lowest lying states, the near trivial
 mixing in the third and fourth ($\chi_{2}$ dominated, $\chi_{1}$
 dominated) and similarly in the fifth and sixth states of the
 $8\times 8$ analysis is manifest.

\begin{figure*}[!t]
  \begin{center}
 
 \includegraphics [height=0.80\textwidth,angle=90]{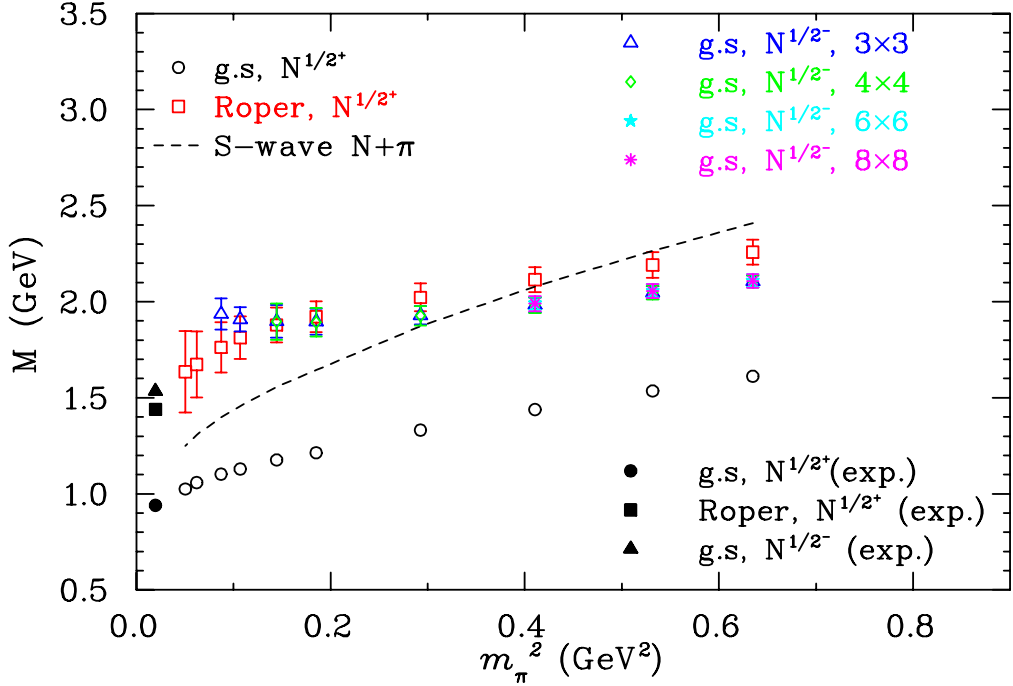}
    \caption{(color online). Mass of the nucleon,
      $N{\frac{1}{2}}^{+}$, Roper $N{\frac{1}{2}}^{+}$ (${\rm
        P}_{11}$) and the
      negative parity ground state, $N{\frac
      {1}{2}}^{-}$ (${\rm{S}}_{11}$) are presented. The combination
      of errors in quadrature is shown. The ground and Roper states
      are taken from Ref.~\cite{Mahbub:2009aa} for comparison with the
      negative parity ground state. The black filled symbols are the
      experimental values.
      } 
   \label{fig:paper_m_ground_Roper_negP_FINAL}
  \end{center}
\end{figure*}

In Fig.~\ref{fig:paper_m_ground_Roper_negP_FINAL}, the negative parity
ground state results are presented for all the correlation
matrices (masses are given in
Table~\ref{table:mpi.mn.gs.3x3.4x4.6x6.8x8}).  The positive parity
ground state (nucleon) and Roper state 
results are taken from Ref.~\cite{Mahbub:2009aa} for
comparison. Regarding the level ordering problem, we do not see the physical
level ordering for the three heavier quark masses, where the
$N{\frac{1}{2}}^{-}$ state lies below the Roper,
$N{\frac{1}{2}}^{+}$, state. A coincidence of the Roper and
$N{\frac{1}{2}}^{-}$ states occurs at pion mass of 380 MeV 
and the physical level ordering is observed at pion masses below 380
MeV.

\input{./table.mpi.mn.gs.3x3.4x4.6x6.8x8.tex}

 For evidence of the physical level ordering, we
 calculate  the correlated jackknife error over the ratio of masses
of Roper to the negative parity state. For our second
lightest quark mass the result is ${0.947}^{\, +0.047}_{\, -0.047}$ and for the
lightest quark mass it is ${0.909}^{\, +0.054}_{\, -0.059}$, providing
strong evidence of a physical level ordering. 
Little evidence is found that our extracted lattice $N{\frac{1}{2}}^{-}$ state
approaches the physical state. We note that chiral physics is
significantly altered  in quenched QCD for $m_{\pi}< 250$
MeV. Moreover, the finite volume of the lattice will also become
significant in this regime. 
Future calculations of negative parity states from the variational
approach  addressing the light quark masses should be done on large
volume  lattices in $2+1$ flavour QCD.

 At our heavier three quark masses the
$N{\frac{1}{2}}^{-}$ state lies below the S-wave $N+\pi$ state
indicative of attractive interactions. The lower values
of the  $N+\pi$ scattering state at light pion masses implies that
the couplings of our interpolators are small for the
 two particle $N+\pi$ state at these pion masses.

\section{Conclusions}
 \label{sec:conclusions}
An extensive analysis has been performed to address the
issue of the level ordering problem in the nucleon sector through the
variational approach. A
similar method to that of Ref.~\cite{Mahbub:2009aa} is used to explore this
long-standing problem. The physical level ordering between the Roper and
$N{\frac{1}{2}}^{-}$ ground states in a pion mass region of 295 -- 380
MeV is observed. However, an identification of a low-lying
negative-parity state in the Physical quark mass regime remains.

\section*{Acknowledgments}
This research was undertaken on the NCI National Facility in Canberra,
Australia, which is supported by the Australian Commonwealth
Government. We also acknowledge eResearch SA for generous
grants of supercomputing time which have enabled this project.  This
research is supported by the Australian Research Council.

\end{document}

%% file: table.mpi.mn.gs.3x3.4x4.6x6.8x8.tex
  \begin{table*}[!t]
    \begin{center}
    \caption{\label{table:mpi.mn.gs.3x3.4x4.6x6.8x8}
      Mass of the negative parity ground state of the nucleon,
      $N{\frac{1}{2}}^{-}$, for $3\times 3$, $4\times 4$, $6\times
      6$ and $8\times 8$ correlation matrices. Masses are averaged over
      the bases described in the text and errors are a combination of
      average statistical errors  over the bases and  systematic
      errors for choice of basis.}
   \vspace{0.5cm}
    \begin{tabular}{ccccc} 
        \hline
        \hline
   $aM_{\pi}$ & $aM^{N^{{\frac{1}{2}}^{-}}}_{\rm{g.s}} \, (3\times 3)$ & $aM^{N^{{\frac{1}{2}}^{-}}}_{\rm{g.s}} \, (4\times 4)$ & $aM^{N^{{\frac{1}{2}}^{-}}}_{\rm{g.s}} \, (6\times 6)$ & $aM^{N^{{\frac{1}{2}}^{-}}}_{\rm{g.s}} \, (8\times 8)$ \\

  \hline 
 0.5141(19) & 1.358(19) & 1.358(20)  & 1.362(21) & 1.359(23) \\   
 0.4705(20) & 1.320(21) & 1.320(22)  & 1.327(22) & 1.326(24) \\
 0.4134(22) & 1.279(25) & 1.278(26)  & 1.285(22) & 1.283(25) \\
 0.3490(24) & 1.244(30) & 1.243(33)  &   -       &   -       \\
 0.2776(24) & 1.223(44) & 1.221(48)  &   -       &   -       \\
 0.2452(24) & 1.224(55) & 1.223(61)  &   -       &   -       \\
 0.2110(27) & 1.231(40) &   -        &   -       &   -       \\
 0.1905(31) & 1.249(53) &   -        &   -       &   -       \\

\hline
\hline
  \end{tabular}
 \end{center}
 \end{table*}